# Thermalization of field driven quantum systems

H. Fotso, K. Mikelsons, and J. K. Freericks[*]

Department of Physics, Georgetown University, 37th and O Sts. NW, Washington, DC 20057 USA

[*]freericks@physics.georgetown.edu (Corresponding author)

October 23, 2013

**There is much interest in how quantum systems thermalize after a sudden change, because unitary evolution should preclude thermalization. The eigenstate thermalization hypothesis resolves this because all observables for quantum states in a small energy window have essentially the same value; it is violated for integrable systems due to the infinite number of conserved quantities. Here, we show that when a system is driven by a DC electric field there are five generic behaviors: (i) monotonic or (ii) oscillatory approach to an infinite-temperature steady state; (iii) monotonic or (iv) oscillatory approach to a nonthermal steady state; or (v) evolution to an oscillatory state. Examining the Hubbard model (which thermalizes under a quench) and the Falicov-Kimball model (which does not), we find both exhibit scenarios (i-iv), while only Hubbard shows scenario (v). This shows richer behavior than in interaction quenches and integrability in the absence of a field plays no role.**

The classical picture for how an isolated system evolves as it is driven by a DC electric field (**E**) is that a current ⟨**j**(t)⟩ develops which subsequently creates heat due to Joule heating at a rate given by ⟨**j**(t)⟩•**E**. Thus the system evolves over time to infinite temperature as the current decreases to zero and ends up in a steady thermal state, with a density of states modified by the field. Such behavior should occur in any interacting system (integrable or not). This then opens the question, do driven quantum systems evolve in a similar way as they do under an interaction quench [1-4]? There are many similarities in these two systems to suggest that they should. A DC field being turned on modifies the Hamiltonian at an instant of time, and for future times, it can be described by a time-independent Hamiltonian in the scalar potential only gauge. If the system thermalizes to the infinite-temperature state when it is field driven, then it also evolves into a steady thermal state in the infinite-time limit.

But there are differences too, the most important being that the field-driven system has a current flowing through it as it thermalizes, so its energy continuously evolves as opposed to the instantaneous change in an interaction quench, and it is much more likely to show oscillatory behavior due to the possibility of Bloch oscillations. For example, driving a noninteracting single-band system with a DC field creates an oscillating current, and the heating varies periodically in time so that the system returns to its initial equilibrium state after each Bloch period given by $2\pi/E$, resulting in no net heating; in particular, it never evolves to the infinite-temperature state. This occurs because there is no scattering in the system that could allow it to thermalize. In this work, we describe what happens for the general case when the field-driven system is interacting.

We consider two interacting quantum systems the Hubbard and the Falicov-Kimball (FK) models, which have been respectively shown to thermalize [5] and not to thermalize [6] under an interaction quench (the FK model has an infinite number of conserved quantities, but is *not* integrable via the Bethe ansatz in one dimension). They are initially in equilibrium and we study their long time behavior after a constant DC electric field is turned on at $t=0$. We then track the real-time transient behavior as the systems evolve toward a steady state and find that the formation of a nonequilibrium steady state density of states (DOS) is only constrained by a "causality" timescale (set by its Fourier transform in the time domain) that plays no further role in the relaxation of the system. At half-filling, an identity relates the DOS to the imaginary part of the lesser Green's function (defined and proven below), so in this case, it is the real part of the lesser Green's function, $\text{Re}[G^<(T_{ave},t_{rel})]$ with $T_{ave}=(t+t')/2$ the average time and $t_{rel}=t-t'$ the relative time, that determines the thermalization. It vanishes for an infinite temperature thermal state and is nonzero for nonequilibrium steady or oscillatory states. We use the evolution of this Green's function (and the current, the kinetic and the potential energies) as a function of time to describe the different scenarios seen for the long-time evolution of the system.

The equilibrium Hamiltonian describing the Hubbard [7] and FK [8] models is

$$H = -\sum_{ij\sigma} J^{hop}_{ij\sigma} c^\dagger_{i\sigma} c_{j\sigma} - \sum_{i\sigma} \mu_\sigma c^\dagger_{i\sigma} c_{i\sigma} + U \sum_i c^\dagger_{i\uparrow} c_{i\uparrow} c^\dagger_{i\downarrow} c_{i\downarrow} \ , \qquad (1)$$

where $c^\dagger_{i\sigma}$ ($c_{i\sigma}$) are the creation (annihilation) operators for an electron at site $i$ with spin $\sigma$, $\mu_\sigma$ is the chemical potential for the corresponding electron (which is independent of spin for the Hubbard model), and $U$ is the interaction energy. The hopping integral is nonzero only between nearest neighbors and serves as our energy unit. We take $J^{hop}_{ij\uparrow} = J^{hop}_{ij\downarrow} = J/(2\sqrt{d})$ for the Hubbard model, where $d$ is the spatial dimension of the lattice; for the FK model, the up spin hopping is the same, but the down spin hopping vanishes since those electrons are localized and do not hop. The hopping $J$ is used as the energy unit.

The nonequilibrium case has a spatially uniform, but time varying, electric field, that is described by a vector-potential-only gauge, where $\boldsymbol{E}(t)=-\partial \boldsymbol{A}(t)/\partial t$ and we have set $c=e=\hbar=1$. The hopping integral becomes time dependent, acquiring a phase, $J^{hop}_{ij\sigma}(t) \to J^{hop}_{ij\sigma}\exp\{-i[\boldsymbol{R}_i - \boldsymbol{R}_j]\cdot \boldsymbol{A}(t)\}$ with $\boldsymbol{R}_i$ the position vector for site $i$ on the lattice [9]. We start the system in equilibrium at an initial temperature, and then instantly turn on a constant electric field, whose spatial component is $E$, at time $t=0$. The field usually points in the diagonal direction $\boldsymbol{E}=(E,E,E,...)$, or in the axial direction $\boldsymbol{E}=(E,0,0,...)$. The FK model is solved exactly on the infinite-dimensional hypercubic lattice, using nonequilibrium dynamical mean-field theory [10,11], while the Hubbard model is solved approximately using a strong-coupling perturbation theory in the hopping on a three-dimensional simple cubic lattice [12].

**Results.** In Fig. 1, we plot the effective temperature, determined by equating the transient energy of the nonequilibrium system at time *t* to the equilibrium energy at temperature *T* in order to obtain the effective temperature *T(t)*. (Note that extraction of an effective temperature by equating the instantaneous nonequilibrium energy to the equilibrium energy at a given temperature does not necessarily imply that the system is in equilibrium at that instant, instead it is a convenient way to keep track of what an effective temperature for the system should be, and as the temperature approaches infinity, the system approaches closer and closer to a true equilibrium state.) As the time approaches infinity, there is a clear evolution toward an infinite-temperature result for cases that thermalize and a clear evolution to a finite temperature (indicating a nonthermal evolution, since the system will not be in equilibrium) for those that do not. The limiting behavior can be approached in a monotonic or oscillatory fashion (with a dynamical phase transition in between [5]). Note that while the results for the FK model, where our algorithm is numerically exact, have a clearer differentiation of the different categories than in the Hubbard model, where our results are approximate, we cannot rule out the possibility that the nonthermal Hubbard model states are prethermalized states that subsequently evolve to the infinite-temperature limit at very long times [13,5,14]. But our results show no indication of this, and such timescales are rarely relevant to any kind of real experiment when the interaction is large enough.

In Fig. 2, we examine the properties of the transient local DOS of the system, which are determined by the local retarded Green's function, after Fourier transformation with respect to the relative time: $A(T_{ave}, \omega) = -\text{Im}\{\int dt_{rel}\ G^R(T_{ave}, t_{rel}) \exp(i\omega t_{rel})/\pi\}$. The DOS typically has a finite bandwidth and no singularities, and hence it is described by a Green function in time that has a finite extent in $t_{rel}$ that we denote as the dynamic range time $t_{dyn}$. This is determined primarily by the inverse of the bandwidth of the density of states (as a function of frequency), as long as the density of states does not have any sharp features or structures that can give rise to long tails in the time domain. For a fixed $T_{ave}$, there always is some range of $t_{rel}$ such that *t'*<0 is before the field is turned on and *t*>0 is after, with the causality line determined by $t_{rel} = 2T_{ave}$. When both times have the field on, the DOS is described by the nonequilibrium steady state DOS. When *t*>0 is after and *t'*<0 before, the DOS is described by the mixed Green's function, interpolating between the nonequilibrium and original equilibrium DOS. Once the causality line passes the dynamic-range time ($2T_{ave} = t_{dyn}$), the transient local DOS is given by the steady state result. This often is a short time for interacting systems, but is infinitely long for the noninteracting single band case, since the steady state DOS is given by the series of delta functions describing the Wannier-Stark ladder, which has an infinite dynamic range time. We illustrate the generic situation in Fig. 2, which shows these scenarios for the Falicov-Kimball and Hubbard models. Only the imaginary part of the retarded Green's function is shown because the real part vanishes at half filling, as explained next.

The retarded Green's function $G_{ij}^R(t, t')$ and the lesser Green's functions $G_{ij}^<(t, t')$ are defined via

$$G_{ij}^R(t,t') = -i\theta(t-t')\text{Tr}e^{-\beta H(t\to-\infty)}\{c_i(t), c_j^\dagger(t')\}_+/Z \tag{2}$$

and

$$G_{ij}^<(t,t') = i\text{Tr}e^{-\beta H(t\to-\infty)}c_j^\dagger(t')c_i(t)/Z \quad, \tag{3}$$

where $\theta(t)$ is the unit step function, $\{.,.\}_+$ denotes the anticommutator, $Z = \text{Tr}e^{-\beta H(t\to-\infty)}$ is the initial (equilibrium) partition function and all operators are expressed in the Heisenberg representation. The local lesser Green's function satisfies a simple identity given by

$$G_{ii}^<(t,t')^* = -G_{ii}^<(t',t), \tag{4}$$

which follows from the definition of the lesser Green's function in Eq. (3) and the invariance of the trace under a cyclic reordering of its terms. If we express this in terms of the Wigner time coordinates, then we find that $\text{Re}G_{ii}^<(T_{ave}, t_{rel}) = -\text{Re}G_{ii}^<(T_{ave}, -t_{rel})$ and $\text{Im}G_{ii}^<(T_{ave}, t_{rel}) = \text{Im}G_{ii}^<(T_{ave}, -t_{rel})$, or the real part of the local lesser Green's function is an odd function of relative time and the imaginary part is an even function of relative time. Next, we examine the retarded Green's function under a particle-hole transformation. Here we assume that the lattice is a bipartite lattice, and the hopping is between the different sublattices only. If the chemical potential satisfies $\mu = U/2$, then the Hamiltonian is invariant under the unitary particle-hole transformation given by

$$c_i^\dagger \to (-1)^{\epsilon(i)}\tilde{c}_i \text{ and } c_i \to (-1)^{\epsilon(i)}\tilde{c}_i^\dagger, \tag{5}$$

where $\epsilon(i)$ is 1 if $i$ is on the A sublattice and 0 if $i$ is on the B sublattice. In the Hubbard model, the particle-hole transformation involves both spins, while for the Falicov-Kimball model, the transformation involves only the conduction electrons, while the localized electrons must be transformed via $w_i \to 1 - w_i$. Since this is a unitary transformation, and the Hamiltonian is unchanged by it, we immediately see that

$$-i\langle c_i(t)c_i^\dagger(t')\rangle = -i\langle \tilde{c}_i^\dagger(t)\tilde{c}_i(t')\rangle = -i\langle c_i^\dagger(t)c_i(t')\rangle \tag{6}$$

where the first equality comes from the unitary particle-hole transformation (the minus signs cancel because the two operators are at the same site). The Hamiltonian used in evaluating the middle expectation value is $\tilde{H}$, and the second equality follows from the fact the Hamiltonian is equal to its particle-hole transformed version at half filling ($H = \tilde{H}$). The retarded Green's function then satisfies

$$\begin{aligned}G_{ii}^R(t,t') &= -i\theta(t-t')[\langle c_i(t)c_i^\dagger(t')\rangle + \langle c_i^\dagger(t')c_i(t)\rangle] \\ &= -i\theta(t-t')[\langle c_i^\dagger(t)c_i(t')\rangle + \langle c_i^\dagger(t')c_i(t)\rangle] \\ &= -i\theta(t-t')[\langle c_i^\dagger(t')c_i(t)\rangle^* + \langle c_i^\dagger(t')c_i(t)\rangle] \\ &= \theta(t-t')[-G_i^<(t,t')^* + G_i^<(t,t')]\end{aligned}$$

where the intermediate steps involve applying particle-hole symmetry, taking the complex conjugate and applying the definition of the lesser Green's function. These results immediately show us that $\text{Re} G_{ii}^R(t,t') = 0$ at half filling and $\text{Im} G_{ii}^R(t,t') = \theta(t-t')\text{Im} G_{ii}^<(t,t')$. Hence, the imaginary part of the local lesser Green's function thermalizes rapidly, just like the retarded Green's function does, and the thermalization of the system is encoded in how the real part of the local lesser Green's function behaves as a function of time (it vanishes in the infinite-temperature limit, but is nonzero in other situations).

We focus on the time dependence of the real part of the lesser Green's function. In Fig. 3, we show examples of the four scenarios that lead to thermalized or nonequilibrium steady states in the long-time limit. The approach to this limit is either monotonic (overdamped), or oscillatory (underdamped), and the crossover between these two regimes has been called a nonequilibrium dynamical phase transition [5] (we primarily use the long-time behavior of the total energy to determine the classification). Each scenario is illustrated with four panels. The top figures in each panel show Re $G^<$ for the Falicov-Kimball model (left) and the Hubbard model (right), while the bottom figures in each panel show the current, kinetic, potential, and total energies (left) for the FK model and (right) for the Hubbard model. Because we cannot evolve these systems out to infinite time, we cannot distinguish between a nonthermal steady state and a transient prethermalized state that subsequently evolves to the infinite-temperature thermalized state. However, we have no evidence for that behavior either (even though it is often assumed to occur in the literature). Once the kinetic energy and current are suppressed to zero, we do not expect any further evolution of the system in time; the nonthermal states arise when the current either vanishes before the infinite-temperature state is reached or it oscillates equally up and down to generate no net heating. Note that the thermalized infinite-temperature cases do not have simple equilibrium analogues, because they have a sharply modified DOS due to the driving by the field [15,16,17,12], so that even though they are described by thermal distribution functions, they cannot be described by equilibrium models without a field.

In Fig. 4, we show the final scenario, evolution to a long-term oscillatory state that does not thermalize or become a steady state. This always occurs for a single-band noninteracting system, and here we found an example of this for the Hubbard model. We did not see such behavior in the FK model, but it might be a strong-coupling phenomenon, and the algorithm used to exactly solve that problem cannot be accurately extended to the regime where the interaction is very large due to the need for too small a discretization size along the contour.

Finally, we show an example of how a system that has an exponential approach to the infinite-temperature limit can often be described by a fluctuation-dissipation theorem-like picture. This is done by first extracting an effective temperature by equating the instantaneous energy with the thermodynamic energy evaluated at equilibrium as a function of the temperature as previously shown in Fig. 1. Next, one takes the nonequilibrium steady state DOS (which can be found from short time transient calculations due to its rapid thermalization, as explained above, or can be solved for exactly using a Floquet-like theory for the Falicov-Kimball model [16,17]). Then one simply forms the quasi-equilibrium lesser Green's function by taking the product of the nonequilibrium steady state DOS with the appropriate Fermi factor for each given average time. We can compare this to the transient lesser Green's function, found by Fourier transforming the relevant lesser Green's function for fixed average time as a function of the relative time. These results are shown in the Figure 5. The agreement between these two different approaches points to a quasi-equilibrium regime when the current is almost zero and that can be described by an effective temperature obtained via energy conservation. This picture is in agreement with the general conjecture found in the context of a one-dimensional system of fermions accelerated by an external field [18].

**Discussion.** We have illustrated that the thermalization problem for a field driven nonequilibrium system is complex, showing either monotonic or oscillatory approach to the thermal state or to a nonequilibrium steady state or evolution to an oscillating, periodic state that does not thermalize (with a dynamic phase transition or crossover separating the different regions). How the system thermalizes (or not) under the quench of the interaction strength does not appear to provide any evidence for how it will evolve when driven by a field, as the two models examined illustrate different behavior under a quench, but similar behavior when driven by a field. These results show that the field-driven thermalization problem has much richer behavior than conventional quench problems, which opens up a new realm for analysis of nonequilibrium behavior. The classification of different scenarios is simplified and clarified by examining the behavior in the time domain rather than the more commonly studied frequency domain.

**Methods:** The strong-coupling perturbation theory for the Hubbard model uses a self-consistent second-order expansion for the self-energy, which includes an infinite class of diagrams, but is truncated [12]. The exact solution for the FK model uses a discretized version of nonequilibrium dynamical mean-field theory which was extrapolated to the continuum limit [10,11]. The Kadanoff-Baym-Keldysh technique is employed to solve for the Green's functions. There are two Green's functions to determine. The retarded Green's function $G_{ij}^R(t,t')$ and the lesser Green's function $G_{ij}^<(t,t')$ defined in Eqs. (2) and (3). The corresponding momentum-dependent Green's functions are formed by taking the spatial Fourier transform, since the Green's functions depend only on the spatial difference of the position variables (due to translational invariance of the system). These Green's functions are calculated with the numerically exact dynamical mean-field theory approach for the Falicov-Kimball model and with the self-consistent second-order strong-coupling expansion for the Hubbard model.

The Green's functions can be employed to calculate the kinetic energy, potential energy, and current in the presence of a driving field. The bandstructure satisfies $\varepsilon(k - A(t)) = -\left(\frac{t^*}{\sqrt{d}}\right)\sum_{i=1}^{d}\cos(k_i - A(t))$ in the limit of $d$ approaching infinity, with $A(t)$ the component of the vector potential, which is nonzero only for positive times. The expectation value for the kinetic energy then becomes

$$E_{kin}(t) = \sum_k \varepsilon(k - A(t))\langle c_k^\dagger(t)c_k(t)\rangle = -i\sum_k \varepsilon(k - A(t))G_k^<(t,t) \tag{8}$$

which is determined from the momentum-dependent lesser Green's function (and is per spin, so must be multiplied by two for the Hubbard model). Similarly, if we define the band velocity via $v(k - A(t)) = -\left(\frac{t^*}{\sqrt{d}}\right)\sum_{i=1}^{d}\sin(k_i - A(t))$ also in the limit of $d$ going to infinity, then the current becomes

$$j(t) = \sum_k v(k - A(t))\langle c_k^\dagger(t)c_k(t)\rangle = -i\sum_k v(k - A(t))G_k^<(t,t) \tag{9}$$

which is also per spin and must be multiplied by two for the Hubbard model. The potential energy is proportional to the average double occupancy and requires the time derivative of the Green's function to evaluate it. It satisfies

$$E_{pot}(t) = \frac{U}{N}\sum_i \langle n_{i\uparrow}(t)n_{i\downarrow}(t)\rangle = \frac{\partial G_{ii}^<(t,t')}{\partial t}\bigg|_{t' \to t} + \frac{3U}{4}\langle n_\uparrow(t) + n_\downarrow(t)\rangle - \frac{U}{2} - E_{kin}(t) \,. \tag{10}$$

**Acknowledgments.** The calculations for the FK model were supported by the National Science Foundation under Grant No. DMR-1006605. The calculations for the Hubbard model were supported by the Air Force Office of Scientific Research under the MURI program grant No. FA9559-09-1-0617. High performance computer resources utilized resources under a challenge grant from the High Performance Modernization program of the Department of Defense. J.K.F. was also supported by the McDevitt bequest at Georgetown.

**Author Contributions**: J. K. F. came up with the initial idea for the work and performed the simulations for the FK model. K. M. performed the calculations for the Hubbard model and discovered the benefits of studying thermalization from the time representation. H. F. performed all the data analysis and the initial draft of the manuscript. All authors participated in the writing and revising of the text.

**Competing Financial Interests:** The author(s) declare no competing financial interests.

**Author Information:** Reprints and permissions information is available at www.nature.com/reprints. The authors have no competing financial interests. Correspondence and requests for materials should be addressed to Jim Freericks at freericks at physics dot georgetown dot edu.

Figure 1.

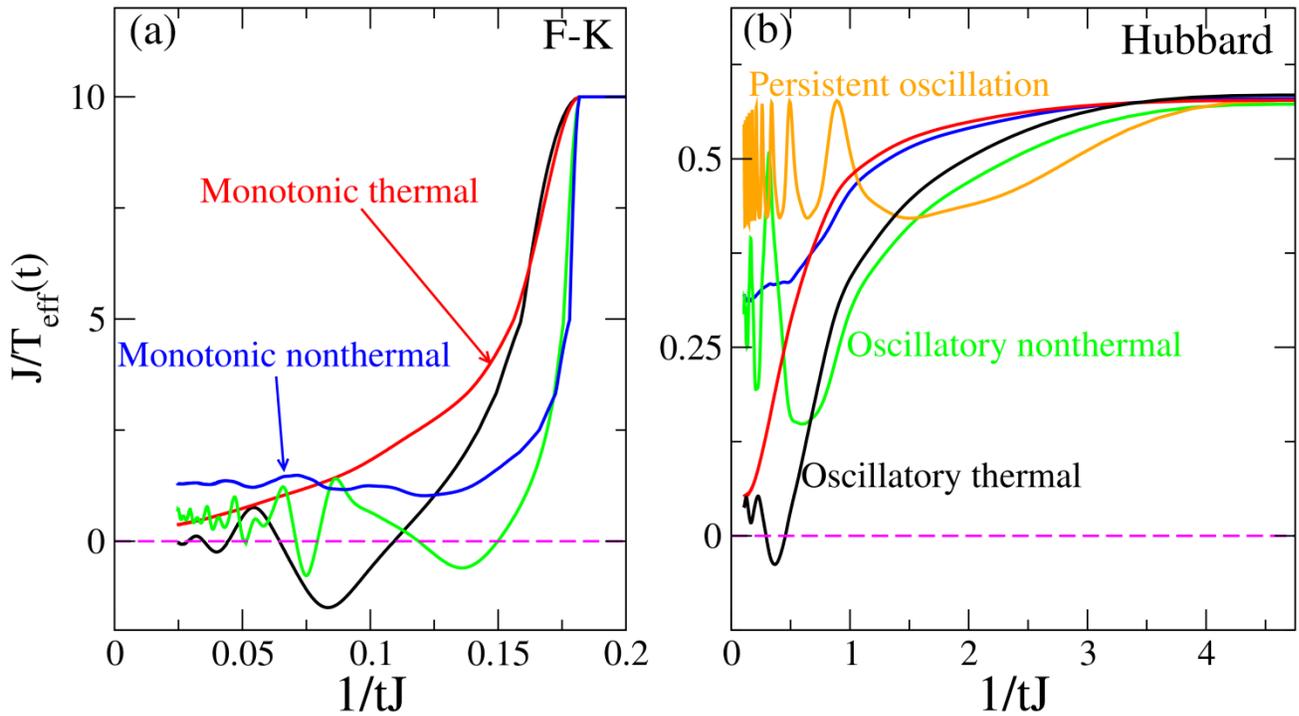

Fig. 1: Effective temperature as a function of time for the half-filled Falicov-Kimball and Hubbard models driven by a field. The colors indicate the different scenarios detailed in the text. The parameters can be found for the corresponding cases in the caption to Figs. 3 and 4. The magenta dashed line indicates the infinite temperature limit corresponding to a thermalized system. While we cannot rule out the possibility of the non-thermal states thermalizing on longer time scales, we see no indication of this occurring in any of the data we analyzed.

Figure 2.

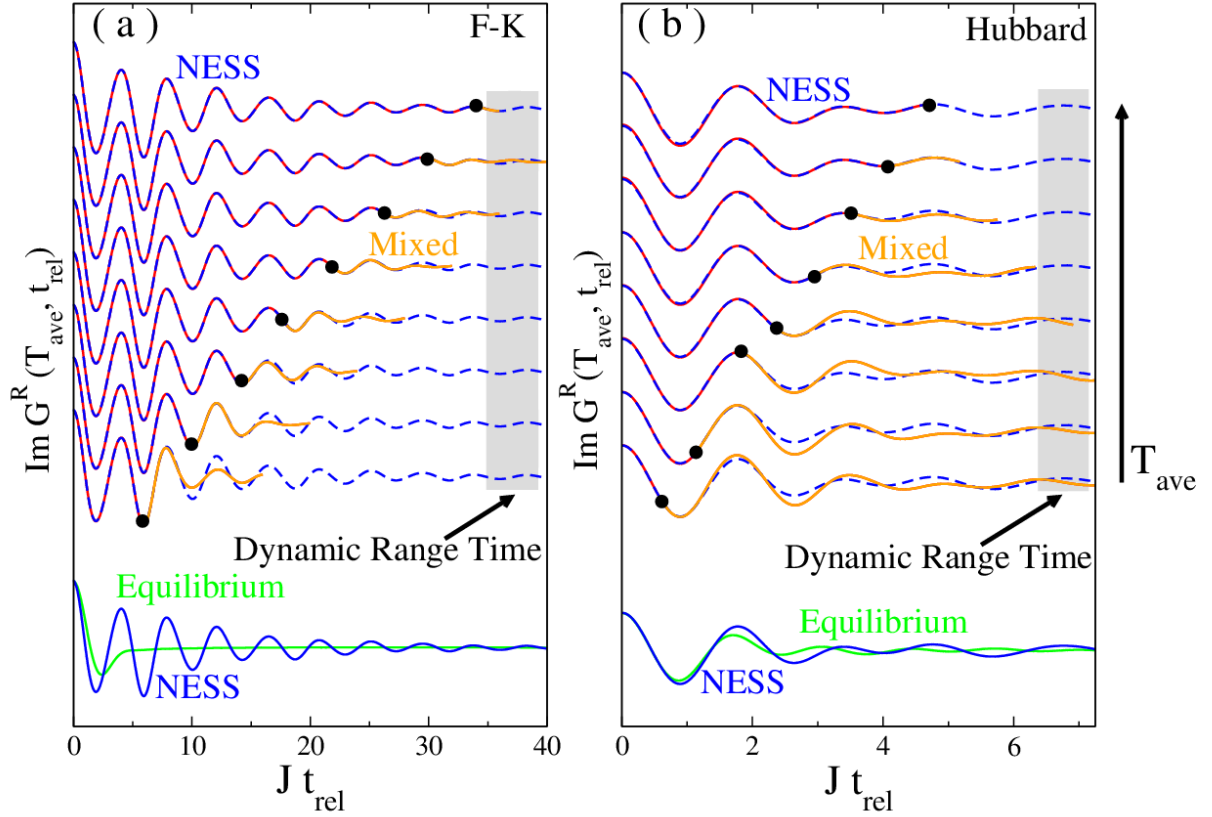

Fig. 2: Imaginary part of the local retarded Green's function as a function of relative time for various average times for (a) the Falicov-Kimball model and (b) the Hubbard model. The lowest curve shows the equilibrium (green) and the nonequilibrium steady state (NESS, blue) $G^R(t_{rel})$; The upper curves show $G^R(T_{ave},t_{rel})$ for successive average times after the field is switched on superposed on the steady state result (blue). Two regions are highlighted: (i) one in which both $t$ and $t'$ have the field on (red, overlapped by the blue) and the other in which $t$ has the field on and $t'$ does not (orange, mixed Green's function). The black dots mark causality time between the two regions. The parameters are as follows: FK model; E=1.0 , U=3.0 , T=0.1, Hubbard model; U=6√3, T=√3, E=(6√3,6√3,6√3). Note how the Green's function becomes negligible in size when we reach the dynamic range time on the right hand side of each panel, indicated approximately by the gray shaded region.

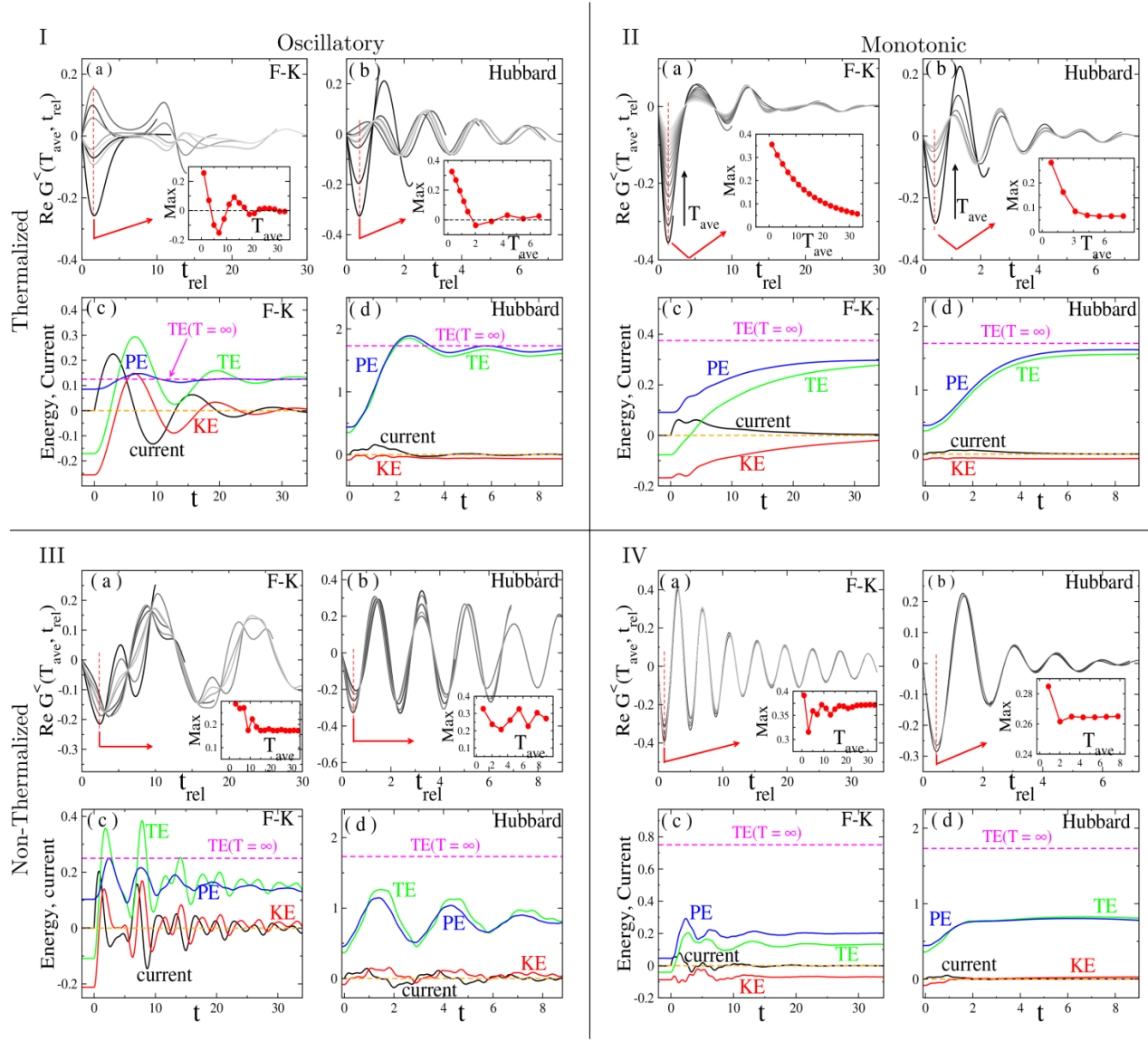

Fig.3. Each of the panels I through IV represents one of the relaxation scenarios. Real part of the lesser Green's function as a function of $t_{rel}$ for various average times, (a) Falicov-Kimball, (b) Hubbard model; A grayscale is used with lighter shades indicating later average times. The inset in both cases shows the amplitude of $G^<(t_{rel})$ as a function of average time. Panels (c) and (d) show for the FK and the Hubbard models respectively, the total energy (green), the potential energy (blue), kinetic energy (red) and the current (black) as a function of time with the same parameters as in (a) and (b), respectively. I (oscillatory, thermalized): FK, E=0.5, U=0.5, T=0.1. Hubbard, U=4√3, E=(1,1,1)x4√3, T=√3. II (monotonic, thermalized): FK, E=0.5, U=1.5, T=0.1. Hubbard, U=4√3, E=(4√3,0,0), T=√3. III (oscillatory, nonthermal): FK, E=2.0, U=1.0, T=0.1. Hubbard, U=4√3, E=(1,1,1)x5√3, T=√3. IV (monotonic, nonthermal): FK, E=2.0, U=3.0, T=0.1. Hubbard, U=4√3, E=(5√3,0,0), T=√3.

Figure 4.

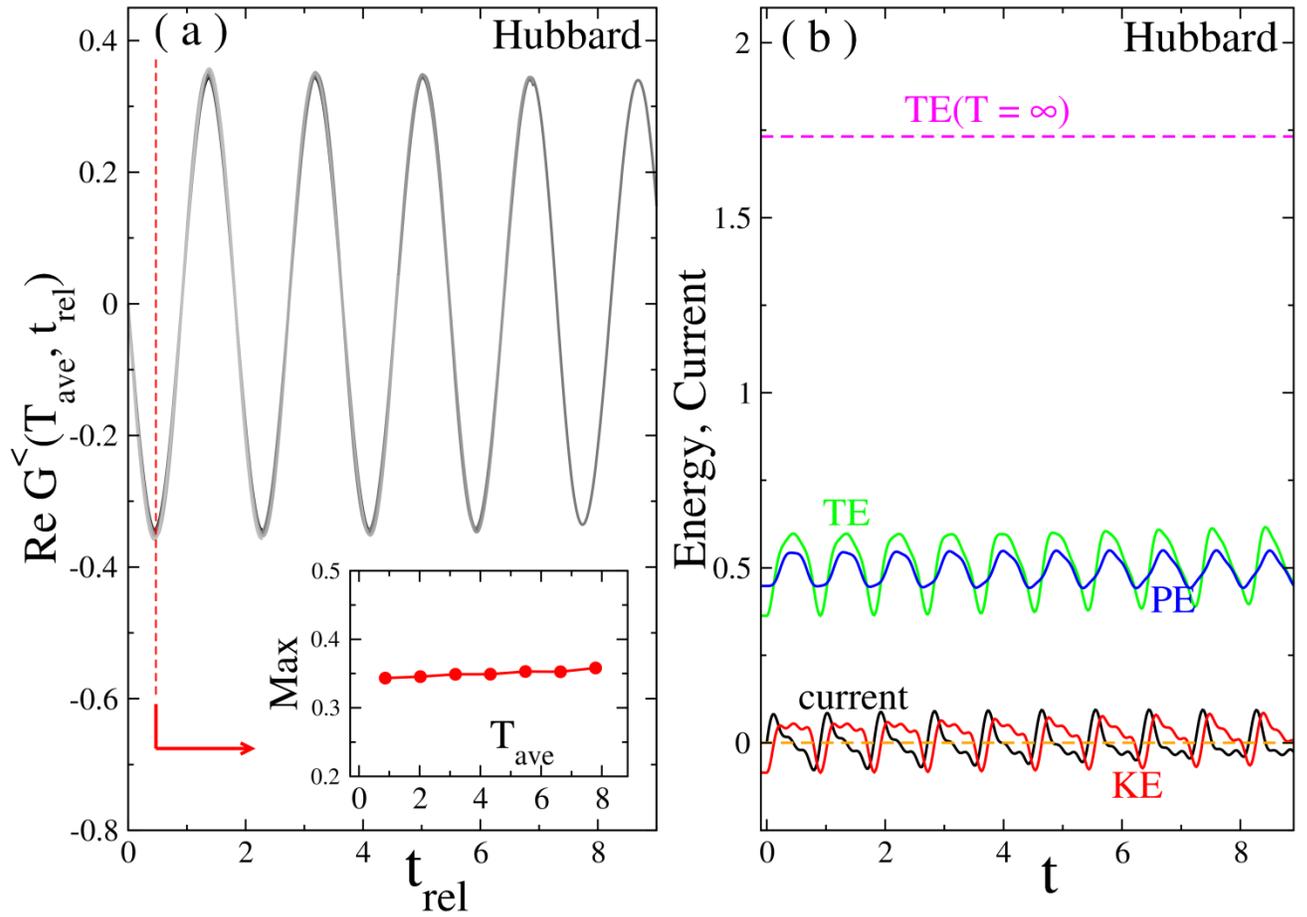

Fig. 4: (a) Lesser Green's function as a function of $t_{rel}$ for various average times for the oscillatory, nondecaying case. A grayscale is used with lighter shades indicating later average times. The inset shows the amplitude of $G^<(t_{rel})$ as a function of average time. (b) shows the total energy (green), the potential energy (blue), kinetic energy (red) and the current (black) as a function of time. Hubbard: $U=4\sqrt{3}$, $E=(2,2,2)\times 4\sqrt{3}$, $T=\sqrt{3}$.

Figure 5.

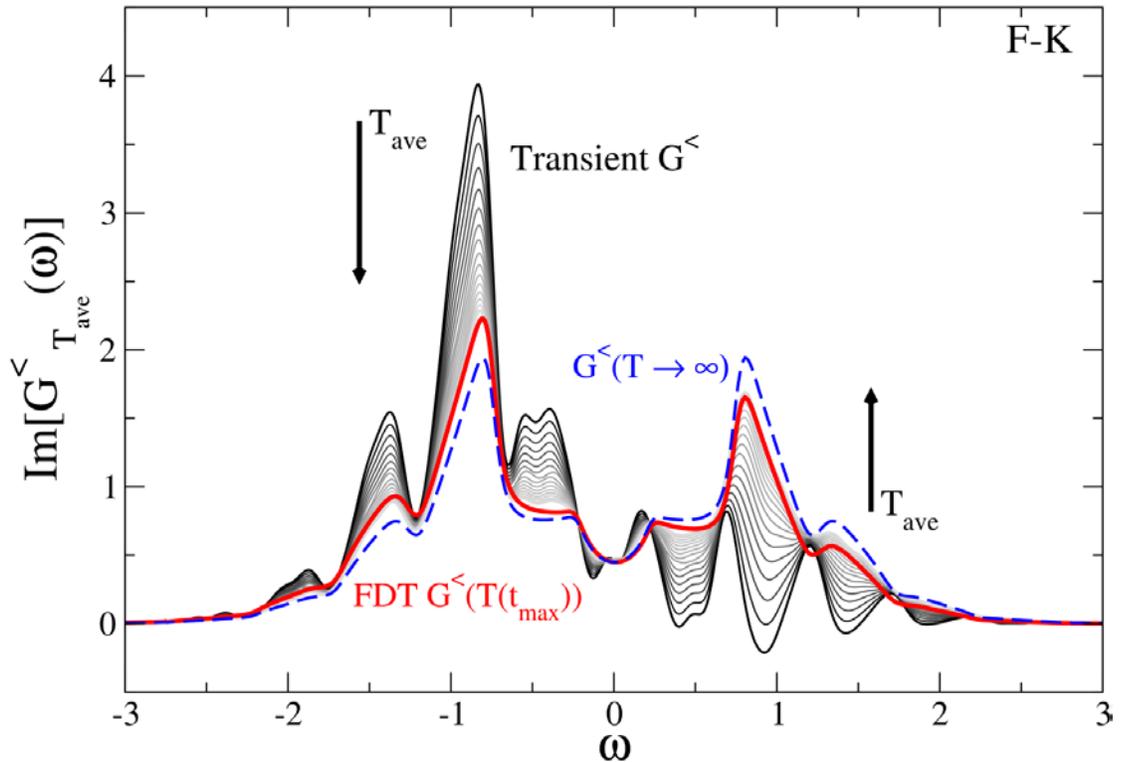

Fig. 5: Imaginary part of the lesser Green's function as a function of frequency for different average times in the Falicov-Kimball model at half filling. As the average time increases, the system converges to the infinite time and infinite temperature limit given by the blue dashed curve. But even before reaching that limit point, the approximation of the lesser Green's function by an appropriate fluctuation-dissipation theorem given by a quasi-equilibrium temperature (red curve, illustrating the longest simulated time) agrees very well with the corresponding transient Green's function extracted directly from the exact solution of the model (series of gray/black curves terminating at the red curve at the longest simulated time). The parameters are $T=0.1$, $U=1.5$, and $E=0.5$, corresponding to a monotonic, thermalized case.